\documentclass{elsart}
\usepackage{graphicx}
\usepackage{dcolumn}
\usepackage{bm}
\usepackage{amsmath}
\usepackage{amsfonts}
\usepackage{amssymb}

\newcommand{\be}{\begin{equation}}
\newcommand{\ee}{\end{equation}}
\newcommand{\bea}{\begin{eqnarray}}
\newcommand{\eea}{\end{eqnarray}}

\newcommand{\ve}{\varepsilon}

\begin{document}
\begin{frontmatter}
\title{Relativistic RPA plus phonon-coupling 
analysis of pygmy dipole resonances}
\author{E. Litvinova}
\address{Physik-Department der Technischen Universit\"at M\"unchen,
Germany, and \\
Institute of Physics and Power Engineering, 249020 Obninsk, Russia}
\author{P. Ring}
\address{Physik-Department der Technischen Universit\"at M\"unchen, 
Germany}
\author{D. Vretenar}
\address{Physics Department, Faculty of Science, University of Zagreb, Croatia, and \\
Physik-Department der Technischen Universit\"at M\"unchen,
Germany}
\date{\today}

\begin{abstract}
The relativistic random-phase approximation (RRPA) plus 
phonon-coupling (PC) model is applied in the analysis of E1 strength 
distributions in $^{208}$Pb and $^{132}$Sn, for which data on 
pygmy dipole resonances (PDR) have recently been reported. 
The covariant response theory is fully consistent:
the effective nuclear interaction NL3 is used 
both to determine the spectrum of single-nucleon 
Dirac states, and as the residual interaction which 
determines the collective phonon states 
in the relativistic RPA.
It is shown that the picture of the PDR as a resonant
oscillation of the neutron skin against the isospin saturated 
proton-neutron core, and with the corresponding RRPA state 
characterized by a coherent superposition of many neutron 
particle-hole configurations, remains essentially unchanged 
when particle-vibration coupling is included. The effect of 
two-phonon admixtures is a weak fragmentation and  
a small shift of PDR states to lower excitation energy.
Even though the PDR calculated in the 
extended model space of $ph \otimes$phonon configurations 
contains sizeable two-phonon admixtures, it basically retains 
a one-phonon character and its dynamics is not modified by
the coupling to low-lying surface vibrations. 

\end{abstract}
\end{frontmatter}

The multipole response of nuclei far from the $\beta$-stability line 
and the possible occurrence of exotic modes of excitation has been 
the subject of a number of recent theoretical and experimental studies.
For neutron-rich nuclei in particular, the pygmy dipole resonance (PDR), 
i.e. the resonant oscillation of the weakly-bound neutron 
skin against the isospin saturated proton-neutron core has 
been investigated. The onset of low-lying E1 strength 
has been observed not only in exotic nuclei with a large neutron 
excess, e.g. for neutron-rich oxygen isotopes \cite{Lei.01}, 
but also in stable nuclei with moderate proton-neutron asymmetry,
like $^{44,48}$Ca and $^{208}$Pb \cite{Rye.02,End.03,Har.04}.
Very recently the dipole strength distribution above the one-neutron 
separation energy was also measured in the unstable $^{130}$Sn and 
the doubly-magic $^{132}$Sn \cite{Adr.05}. In addition to the 
giant dipole resonance (GDR), evidence was reported for a PDR 
structure at excitation energy around 10 MeV both in $^{130}$Sn
and $^{132}$Sn, exhausting a few percent of the E1 
energy-weighted sum rule. 

The interpretation of the dynamics of the 
observed low-lying E1 strength in nuclei
with a pronounced neutron excess is very much under discussion.
Virtually all theoretical analyses, including shell-model 
studies and a number of models based on the random-phase 
approximation (RPA), have shown that in light nuclei, e.g. in 
neutron-rich oxygen isotopes, the low-lying dipole strength is 
not collective and originates from non-resonant single-neutron
excitations. The situation is different in medium-heavy and 
heavy nuclei, where the occurrence of collective PDR states 
has been predicted by several RPA-based calculations, whereas 
other studies, including also RPA-based models, did not find 
collective pygmy states in the energy region below the 
GDR, but only dipole states 
characterized by single-neutron particle-hole configurations.
In particular, studies based on the relativistic RPA 
\cite{Vrepyg1.01,Vrepyg2.01,Paar.03,Paar.05} have shown  
that in neutron-rich nuclei the electric dipole 
response is characterized by the fragmentation of the strength 
distribution and its spreading into the low-energy region.
In contrast to light nuclei where the onset of dipole strength 
in the low-energy region is due to non-resonant single-particle 
excitations of the loosely bound neutrons, in relativistic RPA 
calculations of heavier nuclei low-lying dipole states appear 
which display a more distributed structure of the RPA amplitudes. 
For these nuclei a single collective dipole state is 
identified in the low-energy region, and the characteristic 
dynamics of the pygmy resonance becomes apparent from the analysis of 
the corresponding transition densities and velocity distributions. 
The relativistic RPA analysis of Ref.~\cite{Vrepyg1.01} 
predicted the PDR in $^{208}$Pb at an excitation energy close
to the neutron emission threshold, and subsequently such a
resonance structure was identified in a high-resolution 
($\gamma , \gamma^\prime$) study \cite{Rye.02}, with a 
centroid energy right at the neutron threshold 
(E$_{th} = 7.37$ MeV). In Refs.~\cite{Vrepyg2.01,Paar.03,Paar.05} 
the relativistic RPA and quasiparticle (Q)RPA were employed 
in the analysis of the E1 response in Sn isotopes, and 
the occurrence of the PDR was predicted in neutron-rich Sn nuclei. 
This prediction was confirmed in the recent Coulomb dissociation 
experiment reported in Ref.~\cite{Adr.05}, in which the PDR structure 
was observed in $^{130}$Sn and $^{132}$Sn.

The relativistic RPA and QRPA analyses of the dynamics of low-lying 
E1 strength distributions described above 
were performed on the mean-field level, i.e.
without taking into account the spreading effects which arise from the
coupling of single-nucleon states to the collective low-lying 
excitations (phonons). The principal effect of the particle-vibration 
coupling is an increase of the nucleon effective mass at the Fermi 
surface, and this is reflected in an increase of the density of 
single-nucleon states close to the Fermi energy. It has been argued 
that the inclusion of particle-vibration coupling in (Q)RPA calculations, 
i.e. extending the (Q)RPA model space to include selected 
two-quasiparticle $\otimes$ phonon states, 
would not only improve the agreement 
between the calculated and empirical widths of the GDR structures, 
but it could also have a pronounced effect on the low-lying E1 strength. 
For instance, the coupling to low-lying phonons could fragment the PDR 
structure over a wide region of excitation energies. As a result of 
this fragmentation only an enhancement of the E1 strength would be 
observed in the low-energy region, rather than a prominent PDR peak. 
The importance of particle-vibration coupling effects for the multipole 
response of neutron-rich nuclei has particularly been emphasized 
in studies that have used the QRPA plus phonon coupling model based 
on the Hartree-Fock (Q)RPA with Skyrme effective 
forces \cite{Col.01,Sar.04}. For the neutron-rich oxygen isotopes 
it was shown that the experimentally observed dipole strength 
below 15 MeV \cite{Lei.01} could not be reproduced with a simple 
QRPA calculation, but only with the inclusion of the coupling 
with phonons \cite{Col.01}. In Ref.~\cite{Sar.04} the QRPA plus 
phonon coupling model was applied in the analysis of dipole 
excitations in $^{208}$Pb, $^{120}$Sn and $^{132}$Sn. In 
contrast to the results obtained in the relativistic (Q)RPA 
framework, the QRPA plus phonon coupling model predicts low-lying 
E1 strength of non-collective nature in all three nuclei. 
In particular, from the analysis of the structure of RPA 
amplitudes, it was concluded that none of the the four peaks 
lying below 10 MeV in $^{132}$Sn contains contributions of 
more than two or three different neutron 
particle-hole ($ph$) configurations. 
Predominantly these peaks correspond to just a single-neutron 
transition, and each of them exhausts less than 0.5\% of the 
energy-weighted sum rule. Low-lying E1 excitations in neutron-rich 
Sn isotopes have also been studied in the Quasiparticle Phonon 
Model \cite{TLS.04}, in a model space that included up to 
three-phonon configurations built from a basis of QRPA states, 
and with separable multipole-multipole residual interactions. 
The single-nucleon spectra were calculated for a 
Woods-Saxon potential with adjustable parameters. 
Empirical couplings were
used for the QPM residual interactions. In the QPM spectra for 
$^{120-132}$Sn the low-energy dipole strength was found 
concentrated in a narrow energy interval such that the PDR 
could be identified. A dependence of the PDR strength and 
centroid energies on the neutron-skin thickness was analyzed.
It was shown that, despite significant multi-phonon contributions 
to the mean-energy and transition strength, the PDR states 
basically retain their one-phonon character.

In this work we report the first application of the relativistic 
RPA plus phonon-coupling model in the calculation of the E1 
strength distribution in $^{208}$Pb and $^{132}$Sn. The 
relativistic mean-field framework has recently been extended 
to include the coupling of single-nucleon states to low-lying 
vibrational states (phonons), and its effect on the single-nucleon
spectra has been analyzed \cite{LR.06}. In the present study we 
employ a fully consistent covariant response theory, which 
uses the particle-vibration coupling model in the 
time-blocking approximation (TBA) \cite{Ts.89,KTT.97,Ts.05,LT.05} to 
describe the spreading widths of multipole giant resonances in 
even-even spherical nuclei. In the TBA a special time-projection
technique is used to block the propagation of $ph$ configurations 
through states which have a more complex structure 
than $ph\otimes$phonon. The nuclear response 
can then be explicitly calculated on the $ph\otimes$phonon level 
by summation of infinite series of Feynman's diagrams.

The linear response function is the solution of the 
Bethe-Salpeter equation (BSE) in the particle-hole ($ph$) channel
\be
R(14,23) = {\tilde G}(1,3){\tilde G}(4,2) + \frac{1}{i}\sum\limits_{5678}%
{\tilde G}(1,5){\tilde G}(6,2)W(58,67)R(74,83)\; ,
\label{bse1}
\ee
where the notation for the single-particle quantum numbers includes 
the set of Dirac quantum numbers $\{k_1\}$ and  
the time variable $t_1$: $1 = \{k_1,t_1\}$, and the summation implies also 
integration over the respective time variables.
In addition to the usual particle-hole pairs, the configuration space must 
also include pair-configurations built from positive-energy states occupied 
in the ground-state solution, and empty negative-energy states in the 
Dirac sea \cite{RMG.01}. Thus the set $\{k_i\}$ includes both 
positive- and negative-energy states. The dimension of the configuration 
space is truncated in such a way that the unperturbed particle-hole 
(antiparticle-hole) energies are smaller than $100$ MeV (larger 
than $-1800$ MeV) with respect to the positive-energy continuum. 
The model equations are solved by expanding the nucleon spinors 
in a spherical harmonic oscillator basis \cite{GRT.89}. 
In the present calculation we have used a basis of $20$ oscillator shells.

The amplitude of the $ph$-interaction $W$ in Eq. (\ref{bse1}) reads:
\be
W(14,23) = U(14,23) + i\Sigma^e(1,3){\tilde G}^{-1}(4,2) + 
i{\tilde G}^{-1}(1,3)\Sigma^e(4,2)
- i\Sigma^e(1,3)\Sigma^e(4,2) \; ,
\label{wampl}
\ee
where $\tilde G$ denotes the mean-field single-particle Green's function
\be
{\tilde G}(1,2) = -i\sigma_{k_1}\delta_{k_1k_2}\theta(\sigma_{k_1}\tau)
e^{-i\ve_{k_1}\tau}, \ \ \ \ \ \tau = t_1 - t_2 \; ,
\ee
and $\sigma_{k} = \pm 1$ when $k$ denotes an unoccupied (occupied) state.
$\Sigma^e$ is the energy-dependent part of the relativistic 
mass operator in the Dyson equation for 
the single-particle propagator \cite{LR.06}. 
The origin of this energy dependence is the coupling of single-nucleon 
motion to low-lying collective vibrations,
whose energies and amplitudes are calculated 
with the self-consistent relativistic RPA \cite{RMG.01}.
$U$ is the amplitude of the effective interaction irreducible in
the $ph$ channel. This amplitude is determined as a functional 
derivative of the total nucleon self-energy $\Sigma$
with respect to the exact single-particle Green's function:
\be
U(14,23) = i\frac{\delta\Sigma(4,3)}{\delta G(2,1)}\; ,
\label{uampl}
\ee 
and it can be written as a sum of the 
static mean-field term and a time-dependent term:
\be
U(14,23) = V_{k_1k_4,k_2k_3}\delta(t_{31})\delta(t_{21})\delta(t_{34}) 
+ U^e(14,23),
\ee
where $t_{12} = t_1 - t_2$, and 
\be
U^e(14,23) = i\frac{\delta{\Sigma^e(4,3)}}{\delta G(2,1)} \; .
\label{dcons}
\ee
The details of the solution 
of the Bethe-Salpeter equation (\ref{bse1}) in the 
time-blocking approximation are described in Ref. \cite{RLT.06}.
From the linear response $R(E)$ the strength 
function S(E) is calculated for an external field represented by 
a one-body operator $P$:  
\be
S(E) = - \frac{1}{\pi}\lim\limits_{\Delta\to +0}Im\sum\limits_{k_1k_2k_3k_4}
P_{k_2k_1}R_{k_1k_4,k_2k_3}(E+i\Delta)P_{k_4k_3}^{\ast},
\ee 
where the summation is carried out over the whole Dirac space of 
single-nucleon states, including negative-energy states in the Dirac sea.

The present implementation of the relativistic RPA plus phonon-coupling 
model is fully consistent: the same covariant energy functional 
is used to determine (i) the spectrum of positive- and 
negative-energy single-nucleon states from the self-consistent 
solution of the corresponding system of Dirac 
and Klein-Gordon equations, and (ii) 
to calculate the collective phonon states in the relativistic RPA. 
These two sets of solutions form the basis for the $ph \otimes$phonon
configurations which determine (iii) the particle-phonon coupling amplitudes.  
In the present study we have used the density functional based on 
the standard non-linear effective 
interaction NL3 \cite{NL3} in the calculation of the dipole response
of $^{208}$Pb and $^{132}$Sn. The corresponding RPA phonon spaces 
include collective states with spin and parity: $2^+$, $3^-$, $4^+$, 
$5^-$, and $6^+$, with excitation energies below the neutron 
separation energy $B_n$, and with a reduced transition probability to the 
ground state at least 10\% of the maximal one for a given 
spin and parity. For $^{132}$Sn this criterion includes all the 
phonons $2^+$, $3^-$, $4^+$, $5^-$, and $6^+$ below $B_n$,
whereas in the case of $^{208}$Pb a few very weak modes 
have not been included in the phonon space. 
 
\begin{figure}
\includegraphics[scale=1.30,angle=0]{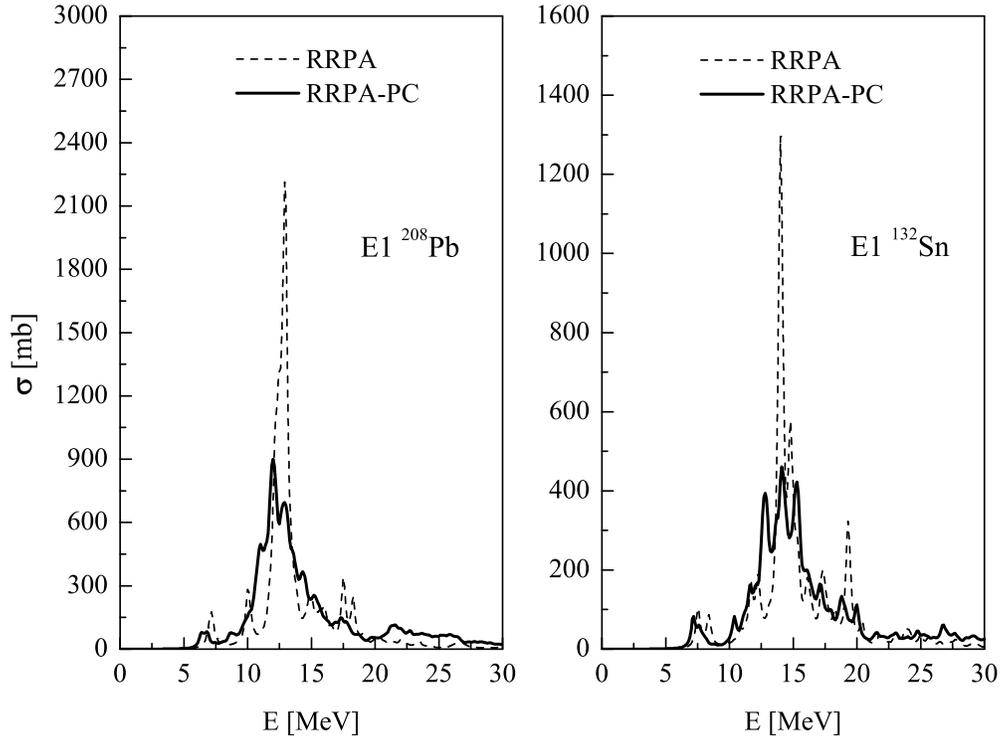}
\caption{E1 photoabsorption cross section for $^{208}$Pb and $^{132}$Sn,
calculated with the relativistic RPA (dashed), and with the RRPA 
extended by the inclusion of particle-phonon coupling (solid).}
\label{fig1}
\end{figure}

The dipole photoabsorption cross sections 
\be
\sigma_{E1}(E) = {{16\pi^3 e^2}\over{9 \hbar c}} E~S_{E1}(E) 
\ee 
for $^{208}$Pb and $^{132}$Sn, calculated  
with the smearing parameter $\Delta = 200$ keV are shown in Fig.~\ref{fig1}.
The corresponding Lorentz fit parameters in the 
two energy intervals: $B_n - 25$ MeV
and $0 - 30$ MeV
are included in Table~\ref{tab1}, in comparison with 
data \cite{Adr.05,ripl}. 
We notice that the inclusion of 
particle-phonon coupling in the RRPA calculation induces a 
pronounced fragmentation of the photoabsorption cross sections, 
and brings 
the width of the GDR in much better agreement with the data, 
both for $^{208}$Pb and $^{132}$Sn. 

\begin{figure}
\includegraphics[scale=0.5,angle=270]{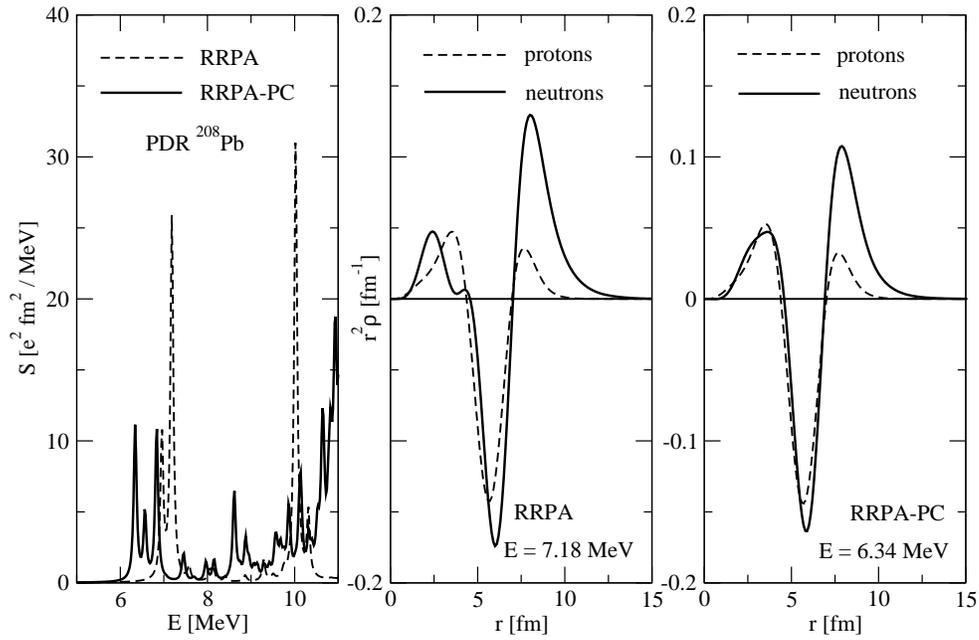}
\caption{The low-energy portion of the E1 strength distribution   
in $^{208}$Pb, calculated with the relativistic RPA (dashed), 
and with the RRPA extended by the inclusion of particle-phonon 
coupling (solid, RRPA-PC). In the panels on the right the proton 
and neutron transition densities for the main peaks below 10 MeV, 
calculated with the RRPA and RRPA-PC, respectively, are plotted as 
functions of the radial coordinate.}
\label{fig2}
\end{figure}

\begin{figure}
\includegraphics[scale=0.5,angle=270]{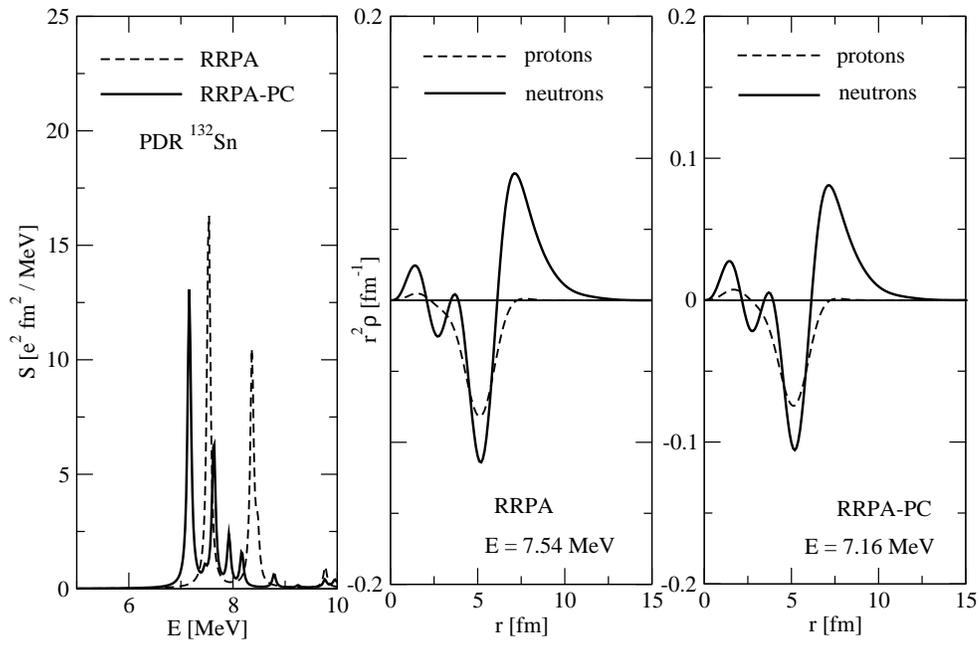}
\caption{Same as in Fig.~\protect\ref{fig2}, but for $^{132}$Sn.
The proton and neutron transition densities correspond to the 
PDR peaks at 7.54 MeV (RRPA) and 7.16 MeV (RRPA-PC).
}
\label{fig3}
\end{figure}

In this work we are more concerned with the effect 
of particle-phonon coupling on the E1 strength function 
in the low-energy region below 10 MeV. The PDR structures 
predicted by our relativistic RPA calculations for $^{208}$Pb
\cite{Vrepyg1.01}, and for $^{132}$Sn \cite{Vrepyg2.01,Paar.03}, 
are clearly visible in the cross sections of Fig.~\ref{fig1}, 
and we also notice that the inclusion of phonon coupling seems to  
have a pronounced effect on these structures. The details are 
shown in Fig.~\ref{fig2} for $^{208}$Pb, and in 
Fig.~\ref{fig3} for $^{132}$Sn, where we display the  
corresponding E1 strength distributions in the low-energy region,
calculated with a smaller value of the smearing parameter
$\Delta = 40$ keV, 
together with the proton and neutron transition densities for the 
strongest peaks below 10 MeV. The PDR peaks calculated with the 
RRPA display characteristic transition densities 
that are very different 
from those of the GDR: the proton and neutron transition densities
are in phase in the nuclear interior, 
there is very little contribution from the protons
in the surface region, the isoscalar transition density dominates
over the isovector one in the interior, and  
the large neutron component in the surface
region contributes to the formation of a node in the isoscalar 
transition density. The low-lying pygmy 
dipole resonance (PDR) does not belong to statistical E1 
excitations sitting on the tail of the GDR, but represents a 
fundamental structure effect: the neutron skin oscillates against the core.
This picture remains essentially unchanged by the inclusion of 
particle-phonon coupling. The principal effect of the coupling with 
phonons in the low-energy region is the redistribution of the E1 
strength and a shift toward lower energies. The main peaks, however, 
retain their basic dynamics, as it can be seen from the proton 
and neutron transition densities. In $^{132}$Sn, in particular, 
the coupling to phonons has the effect of concentrating most of 
the low-lying strength in the PDR peak at 7.16 MeV.

The effect of particle-vibration coupling on the PDR states is also 
illustrated in Table~\ref{tabpb} for $^{208}$Pb, and in 
Table~\ref{tabsn} for $^{132}$Sn, where we display the 
distributions of the neutron particle-hole configurations 
for the most prominent PDR peaks calculated with the RRPA, and 
with the RRPA-PC model. We only include configurations which 
contribute more than 0.1\% to the total RRPA amplitude. 
For the states calculated with the RRPA the percentage 
assigned to a particular $ph$ configuration refers to the 
usual normalization of the RRPA amplitudes for an excited
state $| \nu >$~: 
\be
\sum\limits_{ph}~\Bigl(~ |\rho^{\nu}_{ph}|^2 -
|\rho^{\nu}_{hp}|^2 ~\Bigr) = 1\; .
\ee
We first notice that both in $^{208}$Pb and $^{132}$Sn, many neutron
$ph$ configurations contribute to the RRPA amplitudes of the PDR peaks. 
In $^{208}$Pb we find 14 configurations with more than 0.1\% of the total 
amplitude, with the largest being (3p3/2 $\to$ 3d5/2) with 23.9\%. 
These 14 configurations together contribute 80.2\% to the total 
amplitude, and the remaining $\approx$ 20\% is the contribution of 
proton configurations, and weak neutron $ph$-states with amplitudes
$< 0.1$\%. Note that for GDR states the ratio of neutron to proton 
contribution to the RPA amplitude is typically N/Z, i.e. $\approx 1.5$ 
for $^{208}$Pb, whereas this ratio is more than 4 for the PDR state. An 
extreme case is the PDR at 7.54 MeV in $^{132}$Sn (Table~\ref{tabsn}), 
for which the 10 largest neutron $ph$ amplitudes  
contribute more than 93\% to the total RRPA amplitude. 
The difference between the proton contributions to the PDR in 
$^{208}$Pb and $^{132}$Sn is also seen in the corresponding 
transition densities shown in Figs.~\ref{fig2} and \ref{fig3}.
   
In order to derive the normalization condition for the RRPA plus 
phonon-coupling model, let us rewrite the BS Eq. (\ref{bse1}) by
using the time-projection in the TBA \cite{Ts.89,KTT.97,Ts.05,LT.05},
and performing the Fourier transformation to the energy domain:
\be
R_{k_1k_4,k_2k_3}(\omega) = {\tilde R}_{k_1k_4,k_2k_3}(\omega) - 
\sum\limits_{k_5k_6k_7k_8}
{\tilde R}_{k_1k_6,k_2k_5}(\omega)
\Phi_{k_5k_8,k_6k_7}(\omega) R_{k_7k_4,k_8k_3}(\omega)\; , 
\label{bse}
\ee
where ${\tilde R}$ is the mean-field propagator:
\be
{\tilde R}_{k_1k_4,k_2k_3}(\omega) = -\frac{\sigma_{k_1}\delta_{\sigma_{k_1},-\sigma_{k_2}}
\delta_{k_1k_3}\delta_{k_2k_4}}
{\omega - \ve_{12}}, 
\ee
$\ve_{12} = \ve_{k_1} - \ve_{k_2}$.
$\Phi$ is the generalized amplitude of $ph$ interaction:
\be
\Phi_{k_1k_4,k_2k_3}(\omega) = V_{k_1k_4,k_2k_3} + \Phi^{coupl}_{k_1k_4,k_2k_3}(\omega).
\ee
Close to an eigenfrequency $\Omega^{\nu}$ the response 
function has a simple pole structure:
\be
R^{\nu}_{k_1k_4,k_2k_3}(\omega) = -
\frac{\rho^{\nu}_{k_1k_2}\rho^{{\nu}\ast}_{k_3k_4}}
{\omega - \Omega^{\nu}} \; .
\label{resp}
\ee
One can therefore derive the transition densities for the 
excited state $| \nu >$:
\be
\rho^{\nu}_{k_1k_2} = \lim\limits_{\Delta\to +0}
\sqrt {\frac{\Delta}{\pi S(\Omega^{\nu})}}
\ Im \ \delta\rho_{k_1k_2}(\Omega^{\nu}+i\Delta) ,
\label{ampl}
\ee
where
\be
\quad \delta\rho_{k_1k_2}(\omega) = 
\sum\limits_{k_3k_4} R_{k_1k_4,k_2k_3}(\omega) 
P^{\ast}_{k_4k_3}.
\label{trden}
\ee
It is convenient to rewrite Eq. (\ref{bse}) in the following form:
\be
\sum\limits_{k_5k_6}\Bigl( {\tilde R}_{k_1k_6,k_2k_5}^{-1}(\omega) + \Phi_{k_1k_6,k_2k_5}(\omega) 
\Bigr)R_{k_5k_4,k_6k_3}(\omega) = \delta_{k_1k_3}\delta_{k_2k_4}\delta_{\sigma_{k_1},-\sigma_{k_2}}. 
\label{bsem1}
\ee
By first substituting Eq. (\ref{resp}) into Eq. (\ref{bsem1}), 
and then taking a derivative with respect to
$\omega$, we obtain the generalized normalization condition:
\be
\sum\limits_{k_1k_2k_3k_4}\rho^{{\nu}\ast}_{k_1k_2}
\Bigl[ \sigma_{k_1}\delta_{\sigma_{k_1},-\sigma_{k_2}}\delta_{k_1k_3}\delta_{k_2k_4} - 
\frac{d\Phi_{k_1k_4,k_2k_3}}{d\omega}\mid_{\omega=\Omega^{\nu}} 
\Bigr] \rho^{\nu}_{k_3k_4} = 1 \; , \label{norm}
\ee 
which, in the limiting case of an energy-independent interaction, 
reduces to the usual RPA normalization:
\be
\sum\limits_{k_1k_2}\sigma_{k_1}\delta_{\sigma_{k_1},-\sigma_{k_2}}|\rho^{\nu}_{k_1k_2}|^2 = \sum\limits_{ph}\Bigl( |\rho^{\nu}_{ph}|^2 -
|\rho^{\nu}_{hp}|^2 \Bigr) = 1 \; .
\ee
In the particle-vibration coupling model the derivative in Eq. (\ref{norm})
is a non-positively definite matrix, so that the quantity
\be
\sum\limits_{ph}\Bigl( |\rho^{\nu}_{ph}|^2 - |\rho^{\nu}_{hp}|^2 \Bigr) 
\ee
is always less or equal to 1, analogous to the spectroscopic factor of 
a single-particle state. The difference represents the contribution
of $ph \otimes$phonon configurations.

Because of the coupling to low-lying phonon states, the RRPA pygmy peak at 
7.18 MeV in $^{208}$Pb becomes fragmented and the 
E1 strength is shifted to lower energies. 
In Table~\ref{tabpb} we display the structure of the neutron $ph$ 
configurations for the two most pronounced peaks calculated with 
the RRPA-PC model at 6.84 MeV and 6.34 MeV. 
The contributions of each individual neutron $ph$ configuration 
to the transition amplitude of the state $| \nu >$ are 
quantified by  $|\rho^{\nu}_{ph}|^2 - |\rho^{\nu}_{hp}|^2$ (the 
percentage refers to the generalized normalization of Eq.~(\ref{norm})),
with the amplitudes $\rho^{\nu}_{ph}$ 
and $\rho^{\nu}_{hp}$ calculated from  Eq.~(\ref{ampl}).
12 $ph$ configurations 
contribute with more than 0.1\% to the state at 6.84 MeV, 
compared to 13 for the state at 6.34 MeV. The corresponding 
sums of the amplitudes of the pure neutron $ph$ configurations are now 
reduced with respect to that of the RRPA peak (80.2\%), and 
this reduction indicates the amount of mixing with low-lying phonon
states. The admixture of two-phonon states is especially pronounced 
for the state at 6.34 MeV: only about 50\% of the amplitude corresponds 
to a one-phonon state. Nevertheless, the proton and neutron 
transition densities display the characteristic PDR structure 
(see Fig.~\ref{fig2}). We note that the strong neutron component  
(2f7/2 $\to$ 2g9/2) (14.8\% for the state at 6.84 MeV and 
21.5\% for the state at 6.34 MeV), which is very weak in the 
amplitude of the RRPA pygmy peak at 7.18 MeV (2.5\%), originates 
from the strong RRPA peak at 10.02 MeV. The effect of coupling to 
phonons is much weaker in $^{132}$Sn (see Table~\ref{tabsn}). 
The sum of the amplitudes of the 12 neutron $ph$ configurations 
is reduced by less than 7\% with respect to the RRPA 
calculation. The two-phonon admixture is rather weak, and 
thus the principal effect of coupling with phonon states is
the shift in energy of the PDR state from 7.54 MeV to 7.16 MeV. 

In addition to transitions to bound or quasi-bound states, in  
Tables \ref{tabpb} and \ref{tabsn} we also include transitions 
(with $\geq 0.1$\% of the total amplitude) to single-neutron states 
which belong to the discretized continuum (denoted by the asterisk 
symbol '${\ast}$'). Since their energies depend on the discretization 
scheme (size of the box or, in the present case, the number of oscillator
shells), transitions to these states do not represent physical 
excitations. However, we notice that the contribution of the 
transitions to the discretized continuum is very small, and 
thus our conclusions about the collectivity of pygmy states do 
not depend on the treatment of the continuum. This is, of course, 
to be expected for $^{208}$Pb and $^{132}$Sn, because these nuclei 
are very far from the neutron drip-line and thus threshold effects do 
not play any role in the low-energy multipole response.

The calculated cross sections and the predicted contribution of 
the pygmy resonance to the total dipole strength are compared with 
available data in Table \ref{ics}. We note that the integral cross sections 
$\sigma_{(GDR)}$ calculated in the energy interval $B_n - 25$ MeV are 
in very good agreement with the experimental values, both for 
$^{208}$Pb and $^{132}$Sn. In order to compare the contributions 
of the PDR to the total strength, we have integrated the 
calculated cross sections in the region below 8 MeV (9 MeV) for 
$^{208}$Pb ($^{132}$Sn). The ratios of the resulting $\sigma_{(PDR)}$ 
with $\sigma_{(GDR)}$ are listed in the last column of Table \ref{ics},
and compared with the experimental value 0.03(2) for 
$^{132}$Sn \cite{Adr.05}. The calculated values of this ratio are: 
0.053 for the RRPA, and 0.044 for the RRPA-PC model.

In conclusion, we have applied the relativistic RPA plus 
phonon-coupling model in the analysis of low-lying E1 strength 
distributions in $^{208}$Pb and $^{132}$Sn, for which data on 
pygmy dipole resonances (PDR) have recently been reported. 
The theoretical analysis is fully consistent:
the effective nuclear interaction NL3 is used both
to determine the spectrum of positive- and negative-energy single-nucleon 
Dirac states, and as the residual interaction which 
determines the collective phonon states in the relativistic RPA.
The phonon space to which the single-nucleon states are 
allowed to couple includes phonons with spin and parity:
$2^+$, $3^-$, $4^+$, 
$5^-$, and $6^+$, with excitation energies below the neutron 
separation energy, and with a reduced transition probability to the 
ground state at least 10\% of the maximal one for a given 
spin and parity. The calculated E1 photoabsorption cross sections,  
the excitation energies and widths of the giant dipole resonances 
(GDR) reproduce the available data. In addition the 
RRPA also predicts the occurrence of PDR states in the region of low 
excitation energies below 10 MeV, in agreement with recent 
experimental results. The PDR represents a resonant 
oscillation of the neutron skin against the isospin saturated 
proton-neutron core, and the corresponding RRPA state is 
characterized by a coherent superposition of many neutron 
particle-hole configurations. In this work we have shown that 
this picture remains essentially unchanged when particle-vibration
coupling is included. The effect of two-phonon admixtures is 
a small shift of PDR states to lower excitation energy and, in 
the case of $^{208}$Pb, the PDR strength is fragmented over two 
or three states. Even though the PDR calculated in the 
extended model space of $ph \otimes$phonon configurations 
contains sizeable two-phonon admixtures, it basically retains 
a one-phonon character and its dynamics is not modified by
the coupling to low-lying surface vibrations.

\bigskip 
\leftline{\bf ACKNOWLEDGMENTS}
Helpful discussions with V. Tselyaev are gratefully acknowledged. 
This work has been supported in part by the Bundesministerium f\"ur Bildung
und Forschung under project 06 MT 193, and by the Croatian Ministry of 
Science and Education under project 1191005-1010. E. L. and D.V. acknowledge 
the support of the Alexander von Humboldt-Stiftung.

\bigskip \bigskip

\newpage
\begin{table}[ptb]
\caption{Lorentz fit parameters 
in the two energy intervals: $B_n - 25$ MeV and $0 - 30$ MeV, 
for the E1 photoabsorption cross sections in $^{208}$Pb and $^{132}$Sn, 
calculated with the RRPA, and with the RRPA extended to include
particle-phonon coupling (RRPA-PC), compared to data.}%
\label{tab1}
\begin{center}
\vspace{6mm} 
\begin{tabular}
[c]{cccccccc}
\hline\hline
 & & \multicolumn{3}{c}{$B_n$ - 25 MeV} & \multicolumn{3}{c}{0 - 30 MeV}
\\
 &  & $<$E$>$ & $\Gamma$ & EWSR & $<$E$>$ & $\Gamma$ & EWSR 
 \\
 &  & (MeV) & (MeV) & (\%) & (MeV) & (MeV) & (\%)
\\
\hline
 & RRPA & 13.1 & 2.4 & 121 & 12.9 & 2.0 & 128 
\\
$^{208}$Pb & RRPA-PC & 12.9 & 4.3 & 119 & 13.2 & 3.0 & 128 
\\
 & Exp. \cite{ripl} & 13.4 & 4.1  & 117 & & & 125(8)
\\
\hline
 & RRPA & 14.7 & 3.3 & 116 & 14.5 & 2.6 & 126  
\\
$^{132}$Sn & RRPA-PC & 14.4 & 4.0 & 112 & 14.6 & 3.2 & 126 
\\
 & Exp. \cite{Adr.05} & 16.1(7) & 4.7(2.1)  & 125(32) & & &
\\
\hline\hline
\end{tabular}
\end{center}
\end{table}
\newpage
\begin{table}[ptb]
\caption{Distribution of neutron particle-hole 
configurations for the state at 7.18 MeV (calculated with the RRPA),  
and for the states at 6.84 MeV and 
6.34 MeV (calculated with the RRPA-PC) in $^{208}$Pb. See 
text for the description.}
\label{tabpb}
\begin{center}
\vspace{6mm} 
\begin{tabular}
[c]{ccc}
\hline\hline
  RRPA, 7.18 MeV & RRPA-PC, 6.84 MeV & RRPA-PC, 6.34 MeV \\
\hline
 23.9 \% (3p3/2 $\to$ 3d5/2)   & 40.3 \% (3p3/2 $\to$ 3d5/2) & 21.5 \% (2f7/2  $\to$ 2g9/2) \\
 22.0 \% (3p1/2 $\to$ 3d3/2)   & 14.8 \% (2f7/2 $\to$ 2g9/2) & 14.5 \% (1i13/2 $\to$ 1j15/2) \\
 10.8 \% (1i13/2 $\to$ 1j15/2) &  7.6 \% (3p1/2 $\to$ 3d3/2) & 4.2  \% (3p3/2  $\to$ 3d5/2) \\
 6.6  \% (3p3/2 $\to$ 4s1/2)   &  3.4 \% (2f5/2 $\to$ 3d3/2) & 3.1  \% (3p1/2  $\to$ 3d3/2) \\
 4.2  \% (3p1/2 $\to$ 4s1/2)   &  3.0 \% (3p3/2 $\to$ 4s1/2) & 1.2  \% (3p3/2  $\to$ 4s1/2) \\
 2.9  \% (3p3/2 $\to$ 3d3/2)   &  2.2 \% (2f5/2 $\to$ 3d5/2) & 1.2  \% (1h9/2  $\to$ 1i11/2) \\
 2.6  \% (2f5/2 $\to$ 3d3/2)   &  1.0 \% (3p3/2 $\to$ 3d3/2) & 1.0  \% (1h9/2  $\to$ 2g7/2) \\
 2.5  \% (1h9/2 $\to$ 1i11/2)  &  0.3 \% (1h9/2 $\to$ 1i11/2)& 0.8  \% (3p3/2  $\to$ 3d3/2) \\
 2.5  \% (2f7/2 $\to$ 2g9/2)   &  0.2 \% (2i13/2 $\to$ 1j15/2) & 0.7  \% (2f5/2  $\to$ 2g7/2) \\
 0.7  \% (1h9/2 $\to$ 2g7/2)   &  0.1 \% (2f5/2 $\to$ 4d3/2$^{\ast}$) & 0.3  \% (1h9/2  $\to$ 2g9/2) \\
 0.7  \% (2f5/2 $\to$ 2g7/2)   &  0.1 \% (3p3/2 $\to$ 4d5/2$^{\ast}$) & 0.2  \% (2f7/2  $\to$ 3d5/2) \\
 0.3  \% (2f5/2 $\to$ 3d5/2)   &  0.1 \% (2f7/2 $\to$ 3d5/2) & 0.2  \% (2f5/2  $\to$ 3d3/2) \\
 0.3  \% (2f7/2 $\to$ 3d5/2)   &                             & 0.1  \% (2f7/2  $\to$ 2g7/2) \\
 0.2  \% (1h9/2 $\to$ 2g9/2)   &                             & \                            \\
          80.2  \%             &           73.1 \%           &           49.0 \%            \\
\hline\hline
\end{tabular}  
\end{center}
\end{table}
\newpage
\begin{table}[ptb]
\caption{Same as in Table~\protect\ref{tabpb}, but for 
the states at 7.54 MeV (calculated with the RRPA),  
and at 7.16 MeV (calculated with the RRPA-PC) in $^{132}$Sn. }
\label{tabsn}
\begin{center}
\vspace{6mm} 
\begin{tabular}
[c]{cc}
\hline\hline
  RRPA, 7.54 MeV & RRPA-PC, 7.16 MeV \\
\hline
53.6 \% (3s1/2 $\to$ 3p3/2) & 49.5 \% (3s1/2  $\to$ 3p3/2) \\
16.5 \% (3s1/2 $\to$ 3p1/2) & 21.5 \% (3s1/2  $\to$ 3p1/2) \\
9.7  \% (2d3/2 $\to$ 3p1/2) & 6.4  \% (2d3/2  $\to$ 3p1/2)\\
7.3  \% (2d3/2 $\to$ 3p3/2) & 4.1  \% (1h11/2 $\to$ 1i13/2) \\
4.7  \% (1h11/2 $\to$ 1i13/2)&3.9  \% (2d3/2  $\to$ 3p3/2) \\
0.9  \% (1g7/2 $\to$ 1h9/2) & 0.7  \% (1g7/2  $\to$ 1h9/2) \\
0.3  \% (2d5/2 $\to$ 3p3/2) & 0.1  \% (1g7/2  $\to$ 2f5/2) \\
0.2  \% (1g7/2 $\to$ 2f5/2) & 0.1  \% (2d5/2  $\to$ 3p3/2) \\
0.1  \% (1g7/2 $\to$ 2f7/2) & 0.1  \% (2d3/2  $\to$ 4p1/2$^{\ast}$) \\
0.1  \% (2d3/2 $\to$ 4p1/2$^{\ast}$) & 0.1  \% (1g7/2  $\to$ 2f7/2) \\
     \                      & 0.1  \% (3s1/2  $\to$ 4p1/2$^{\ast}$) \\ 
     \                      & 0.1  \% (3s1/2  $\to$ 4p3/2$^{\ast}$) \\ 
93.4 \%  & 86.7 \% 
\\
\hline\hline      
\end{tabular}
\end{center}
\end{table}
\newpage
\newpage
\newpage
\newpage
\begin{table}[ptb]
\caption{Integral photoabsorption cross sections for the PDR and GDR,
and their ratios calculated with the RRPA and RRPA-PC, 
in comparison with the experimental values. See text for 
description.}%
\label{ics}
\begin{center}
\vspace{6mm} 
\begin{tabular}
[c]{ccccc}
\hline\hline
 & & $\sigma_{(PDR)}$ & $\sigma_{(GDR)}$ & $\sigma_{(PDR)}/\sigma_{(GDR)}$
\\
 & & (mb MeV) & (mb MeV) & 
\\
\hline
 & RRPA & 133 & 3606 & 0.037 
\\
$^{208}$Pb & RRPA-PC & 106 & 3547 & 0.030 
\\
 & Exp. \cite{ripl} &  & 3487  & 
\\
\hline
 & RRPA & 115 & 2162 & 0.053  
\\
$^{132}$Sn & RRPA-PC & 91 & 2087 & 0.044 
\\
 & Exp. \cite{Adr.05} & 75(57) & 2330(590) & 0.03(2)
\\
\hline\hline
\end{tabular}
\end{center}
\end{table}
\end{document}